\documentclass[10pt,a4paper]{article}
\usepackage{pifont}
\usepackage{exscale}
\usepackage[psamsfonts]{amssymb}
\usepackage{amsmath}
\usepackage{epsfig}
\def\lla{\langle\langle}
\def\rla{\rangle\rangle}
\begin{document}
\title{{\bf Magnons in the ferromagnetic Kondo-lattice model}}
\author{M.~Vogt\footnote{Present address: Theory of Condensed Matter Group, 
Cavendish Laboratory, Madingley Road, Cambridge CB3 0HE, UK}, C.~Santos and 
W.~Nolting}
\date{{\em Department of Physics, Humboldt-University at Berlin,\\ 
Invalidenstr. 115, 10115 Berlin, Germany}}
\maketitle
\begin{abstract}
 The magnetic properties of the ferromagnetic
  Kondo-lattice model (FKLM) are investigated. Starting
  from an analysis of the magnon spectrum in the spin-wave regime, we examine 
  the ferromagnetic stability as a function of the
  occupation of the conduction band $n$ and the strength $J$ of the coupling
  between the localised moments and the conduction electrons. From the
  properties of the spin-wave stiffness $D$ the ferromagnetic phase at
  zero temperature is derived. Using an approximate formula the
  critical temperature $T_c$ is calculated as a function of $J$ and $n$.
\end{abstract}

\section{Introduction}
The intensively investigated Kondo-lattice model (KLM) describes the exchange 
coupling of the itinerant electrons to a system of permanent, localised
magnetic moments. These moments have their origin in the partially filled 
5{\textit f}-, 4{\textit f}- and 3{\textit d}-shells of materials such as  
actinides, rare earth elements, transition metals and their compounds. \\
The electrons of the conduction band mediate an indirect exchange
interaction between the local moments, which order collectively below a 
critical temperature $T_c$. The Hamiltonian of the Kondo-lattice model is 
defined as follows \cite{Zene51}:

\begin{equation}
\label{Hsf1}
H=\sum_{ij\sigma} T_{ij} a_{i\sigma}^+a_{j\sigma} -J\sum_{i}{\bf
  \sigma}_i{\bf S}_i.
\end{equation} 
The first term describes the hopping of the conduction electrons between 
different lattice sites at ${\bf R}_i$ and ${\bf R}_j$. $T_{ij}$ are the
hopping matrix elements, which represent the amplitude of the
respective hopping process. The hopping integral is

\begin{equation}
\label{hoptrans}
T_{ij}=\frac{1}{N}\sum_{\bf k}\epsilon({\bf k}) e^{i{\bf k}({\bf R}_i-{\bf
R}_j)},\         
\end{equation}
where $\epsilon({\bf k})$ is the Bloch dispersion.
The operators $a_{i\sigma}^+$ and $a_{j\sigma}$ denote the
creation and annihilation operators for conduction electrons with spin $\sigma$.\\
The second term describes the exchange coupling between the system of 
local moments and the conduction electrons. This term is similar to the
Heisenberg Hamiltonian \cite{Hei28} and describes
the local
character of the interaction. ${\bf \sigma}_i$ is the spin operator of the
conduction electrons, and ${\bf S}_i$ the operator of the localised spins.\\
A very important fact is the sign of the coupling constant 
$J$. If $J<0$, the itinerant
electrons and the local spins couple antiferromagnetically. One  
result is the existence of a nonmagnetic phase caused by screening of
the local moments \cite{Lacr79,Lacr85}. The term ``Kondo-lattice model'' 
implies a negative $J$. If $J>0$ the two subsystems couple ferromagnetically. 
Because the model with 
positive $J$ differs from the KLM only in the sign of $J$ it is refered to as
the ``ferromagnetic Kondo-lattice model (FKLM)'' or ``{\textit{sf-(sd-)}}
model''.\\
Interest in the FKLM has been on the increase following the rediscovery 
of `colossal magnetoresistance' in various manganites, where a large change
in resistivity associated with the FM transition occurs.\cite{Ram97}. The 
first observation of this effect has been made in 1950 \cite{Jonk50}  
Manganites have a 
perovskite structure R$_{1-x}$X$_x$MnO$_3$, where R=La, Pr, Nd and X=Sr, Ca, 
Br, Pb. 
The manganite La$_{1-x}$Ca$_x$MnO$_3$ is a prototype for the double exchange 
model \cite{quantmag2}, which can qualitatively describe ferromagnetism in 
manganites. However,  it cannot reproduce the complex phase   
diagram of La$_{1-x}$Ca$_x$MnO$_3$, where antiferromagnetic, ferromagnetic 
and paramagnetic phases as well as phase separation has been
observed \cite{Ram97}.
\\
A procedure to reproduce this phase diagram using the FKLM has been
presented in a paper by Dagotto {\textit{et al.}\cite{Dago98} These authors
calculate phase diagrams for one- and two-dimensional systems with different 
total spins $S$.   
Both the existence of different magnetic phases and  
phase separation were demonstrated.  A similar phase diagram has been 
calculated for infinite dimensions \cite{Nagai99}. It is generally believed, 
therefore, that the FKLM is a good
starting point for the determination of a phase diagram of manganites even 
in three dimensions. This is plausible if the $t_{2g}$ electrons of the
Mn$^{3+}$ ions in La$_{1-x}$Ca$_x$MnO$_3$  are
assumed to be quasilocalised $S=\frac{3}{2}$-spins and the $e_g$
electrons are assumed to be itinerant electrons moving through the 
$S=\frac{3}{2}$ lattice and being ferromagnetically
coupled to the localised moments.\\
Another class of substances which are described by the FKLM are 
the ferromagnetic metals Gd, Tb, Dy and doped EuX \cite{Proc98}. Gd is 
an excellent prototype since its total spin of $\frac{7}{2}$ is the largest 
possible for the rare earth elements. The main focus of this paper are the
ferromagnetic metals. In the framework of the FKLM the main 
difference between these materials and the manganites is the value of
of the couplings strength $J$, which is one order of magnitude larger
in manganites. The coupling constant in ferromagnetic 
metals is in the order of 0.1 eV \cite{Rex96}.\\      
The main aim of this work is to determine the ferromagnetic phase in
three dimensions at $T=0$ K by analysing the stability of the spin-wave 
spectra. The magnon energies will be calculated  by determining the poles of a
magnon Green's function. The same approach has been used by Wang 
\cite{Wang97}, 
whose numerical results showed two significant properties of 
spin-wave spectra in the FKLM, namely the vanishing of the acoustic
mode in the continuum and the effect of an anomalous
softening of the spin-wave dispersion at the edge of the Brillouin zone.
The second effect has been experimentally demonstrated in the manganite
Pr$_{0.63}$Sr$_{0.37}$MnO$_3$ \cite{Hwang98}. The author of \cite{Wang97}
did not, however, investigate systematically the spin-wave dependence on the
coupling strength and the band occupation $n$. The dependence of the spin
wave energy on the coupling strength in the FKLM has been calculated by 
Furukawa for a fixed band occupation. An increase of the magnon energy with
increasing coupling strength has been shown \cite{Furu96}.\\
In this work a systematic investigation of the spin wave energy at different 
band occupations and coupling strengths is presented. From the spin 
wave stiffness $D$ we can
determine the ferromagnetic phase in a $J-n$ diagram in 3D at $T=0$ K. It 
will 
be of particular interest to see whether and at what parameter configuration 
anomalous softening occurs. Finally we calculate the critical temperature 
dependence on $J$ and $n$ using an approximate formula taken from the 
Heisenberg model.

\section{Theory}

In this section we derive the magnon Green's function for low temperatures in 
the random phase approximation (RPA). We then discuss briefly the resulting 
energetic structure of the magnon system.\\ 
We use the identities for the electron spin operators ($\hbar$=1)\\

\begin{equation}
\sigma_{i}^z=\frac{1}{2}\sum_{\sigma}z_{\sigma}n_{i\sigma},
\sigma_{i}^+=a_{i\uparrow}^+a_{i\downarrow},
\sigma_{i}^-=a_{i\downarrow}^+a_{i\uparrow},
\end{equation}
and for the local moment spin operators

\begin{equation}
\label{ortsst}
S_i^x=\frac{1}{2}(S_i^++S_i^-);\: S_i^y=\frac{1}{2i}(S_i^+-S_i^-)
\end{equation}
to find a more useful form for the {\textit{sf}}-term in Eq.(\ref{Hsf1}) 

\begin{equation}
\label{Hsf2}
H_{sf}=-\frac{J}{2} \sum_{i,\sigma}(z_{\sigma} S_i^z 
n_{i \sigma}+S^{\sigma}_i a_{i-\sigma}^+a_{i \sigma})
\end{equation}  
with $\sigma=\uparrow=+, \sigma=\downarrow=-, z_{\uparrow}=1$ and 
$z_{\downarrow}=-1$. The $S_i^{\sigma}$ are the spin flip operators
which cause spin deviation in the local moment system. \\
The first term on the right-hand side of Eq.(\ref{Hsf2}) is an 
"Ising-term", since it describes the interaction of the 
{\textit{z}}-components of localised spins \cite{Isi25}. The second term is 
non-diagonal in the spin indices and is called the ``spin-flip term''. \\    
We want to investigate the properties of our system at very low
temperatures. In this limit we can apply the spin-wave
approximation using the Holstein-Primakoff transformation \cite{Holst40} . 
This expression can then be simplified for low temperatures. By applying this 
approximation to (\ref{Hsf1}) one gets:

\begin{equation}
\label{hamsf}
H=\sum_{ij} T_{ij}a^+_{i \sigma}a_{j \sigma}-
\frac{J}{2}\left(\sum_{i,\sigma}z_{\sigma}(S-n_i)n_{i\sigma}+
\sum_i\sqrt{2S}(b_i a_{i \downarrow}^+a_{i \uparrow}
+b_i^+ a_{i \uparrow}^+a_{i \downarrow})\right).
\end{equation}
$b_i^+$,$b_i$ are Bose operators.  
The elementary excitation of the spin system can be derived from the retarded
magnon Green's function $\langle\langle S_i^+;S_j^-\rangle\rangle_{E}$, 
which is defined as follows:
 
\begin{equation}
\langle\langle S_i^+;S_j^-\rangle\rangle_{E}=
-{\rm i}\int_{0}^{\infty}\ {\rm d}t\: exp(-{\rm i}Et)\:
\langle [S_i^+(t),S_j^-(0)]_{-} \rangle. 
\end{equation} 
$\langle ... \rangle$ means thermodynamic averaging, while $[...]_{-}$ denotes
the commutator.\\ 
In the spin wave approximation, the magnon Green's function is given by
\begin{equation}
\label{green}
\langle\langle S_i^+;S_j^-\rangle\rangle_{E}=2S \lla b_i;b_j^+\rla_{E}. 
\end{equation}
 The equation of motion of $\lla b_i;b_j^+\rla_{E}$ is

\begin{eqnarray}
\nonumber
E\lla b_i;b_j^+\rla_{E}&=&\langle [b_i,b_j^+]_{-}\rangle +\lla [b_i,H]_-;b_j^+
\rla_{E}
\\
\label{eom1}
&=&\delta_{ij} 
+\frac{1}{2}J\sum_{\sigma}z_{\sigma}\lla b_{i}n_{i\sigma};b_j^+\rla_{E} 
-\frac{1}{2}J\sqrt{2S}\lla a_{i\uparrow}^+a_{i\downarrow};b_j^+\rla_{E}.
\end{eqnarray} 
On the right-hand side of the equation of motion we have two higher 
Green's functions. The spin-flip Green's function represents a characteristic 
part of the
interaction between electrons and magnons since a magnon is
created or annihilated by an electron spin-flip process. These processes are 
of primary importance in determining the magnetic properties of the FKLM.
We therefore decouple the Ising-function only: 

\begin{equation}
\label{ising}
\lla b_i n_{i\sigma};b_j^+\rla_{E} =\langle n_{i\sigma}\rangle \lla 
b_i;b_j^+\rla_{E}. 
\end{equation} 
The spin-flip function is treated more carefully. We define a general
spin-flip Green's function $\Gamma_{ik,j}(E)$:
\begin{equation}
\label{Green2}
\Gamma_{ik,j} (E)=\lla a_{i\uparrow}^+ a_{k\downarrow};b_j^+ \rla_{E}. 
\end{equation}
If we now apply Eq.(\ref{ising}) to the equation of motion (\ref{eom1}) and 
use the Fourier transform $G_{\bf q}(E)=\lla b_{\bf q};b_{\bf q}^+ \rla_{E}$ 
of $G_{ij}(E)=\lla b_i;b_j^+ \rla_{E}$ we get the following form of the 
equation of motion in {\bf q}-space: 

\begin{eqnarray}
\label{eomq}
E G_{\bf q}(E)&=&1+J\langle \sigma^z \rangle G_{\bf q}(E)
-\frac{J}{2N}\sqrt{2S}\sum_{i,j}e^{-i{\bf q}({\bf R}_i-{\bf R}_j)}
\Gamma_{ii,j}(E)\\
\langle \sigma^z \rangle&=&\frac{1}{2N}\sum_{{\bf k},\sigma}z_{\sigma}\langle 
n_{{\bf k}\sigma}\rangle.
\end{eqnarray}
We now derive the equation of motion for $\Gamma_{ik,j}(E)$:

\begin{eqnarray}
\label{eom2}
\nonumber
E\Gamma_{ik,j}(E)&=&\langle
[a_{i\uparrow}a_{k\downarrow},b_j^+]_{-}\rangle+
\lla [a_{i\uparrow}a_{k\downarrow},H]_{-};b_j^+\rla_{E} \\\nonumber
&=&\sum_n(T_{kn} \Gamma_{in,j}(E)-T_{ni}\Gamma_{nk,j}(E))
+JS \Gamma_{ik,j}(E) \\
&-&  \frac{J}{2}\left( \lla n_k a_{i\uparrow}^+a_{k\downarrow};b_j^+\rla_{E}
+\lla n_i a_{i\uparrow}^+a_{k\downarrow};b_j^+\rla_{E}\right) \\
&-&\frac{J}{2}\sqrt{2S}\left(\lla b_k a_{i\uparrow}^+a_{k\uparrow};b_j^+\rla_{E}
-\lla b_i a_{i\downarrow}^+a_{k\downarrow};b_j^+\rla_{E}\right).\nonumber
\end{eqnarray} 
On the right-hand side of Eq.(\ref{eom2}) we have a set of higher Green's
function which describe more complex interaction processes between
itinerant electrons and magnons. These higher Green's functions are 
RPA decoupled, preserving spin and particle conservation, as follows: 

\begin{eqnarray}
\label{dec}
\nonumber
\lla n_k a_{i\uparrow}^+a_{k\downarrow};b_j^+\rla_{E} &\approx&\langle n_k \rangle
\Gamma_{ik,j}(E)\\
\lla b_k a_{i\uparrow}^+a_{k\uparrow};b_j^+\rla_{E} &\approx&\langle 
a_{i\uparrow}^+a_{k\uparrow} \rangle G_{kj}(E). 
\end{eqnarray} 
Applying the Fourier transform of the spin-flip Green's function to
\ref{eom2} we
and using Eq.(\ref{hoptrans}) we obtain the following approximate expression 
for the equation of motion: 

\begin{equation}
\label{gammaq}
\Gamma_{{\bf k,q'}}(E)=-\frac{J\sqrt{2S}}{2\sqrt{N}}
\frac{(\langle n_{{\bf k}\uparrow}\rangle-\langle n_{{\bf k+q'}\downarrow}
\rangle)\:G_{\bf q'}(E)} 
{E-\epsilon({\bf k+q'})+\epsilon({\bf k})-J(S-\frac{1}{N}\sum_{\bf p} 
\langle n_{\bf p} \rangle)}.  
\end{equation}  
At this stage we can obtain an analytical expression of the magnon Green's 
function by solving Eq.(\ref{eomq}):

\begin{equation}
\label{Gq}
G_{\bf q}(E)=\frac{1}{E-J\langle \sigma^z \rangle -J^2S\chi({\bf q},E)}
\end{equation}
where

\begin{equation}
\label{chi}
\chi({\bf q},E)=\frac{1}{2N}
\sum_{\bf k}\frac{\langle n_{{\bf k}\uparrow}\rangle-\langle n_{{\bf k+q}\downarrow}
\rangle} 
{E-\epsilon({\bf k+q})+\epsilon({\bf k})-J(S-\frac{1}{N}\sum_{\bf p} 
\langle n_{\bf p} \rangle)}
\end{equation}
and $\langle n_{\bf p} \rangle$ is the occupation number of the magnons. 
At very low 
temperatures the number of magnons in the system is very small since the 
system is almost saturated ferromagnetically. We therefore set $\langle 
n_{\bf p}\rangle=0$ for numerical simplicity. \\
In \cite{Wang97} the magnon Green's function $\lla S^+(-{\bf q});S^-({\bf q}) 
\rla$ is derived by solving the equation of motion for finite temperatures in 
RPA. 
If we set $\langle S^z \rangle=S$ in this more general expression, our result 
with
$\langle n_{\bf p}\rangle=0$ and ${\bf q}=-{\bf q}$ is obtained. \\
In the RPA theory, the magnon energy is renormalised by the term 
$\chi({\bf q},E)$, which is due to electron-magnon interaction. This term is 
given explicitly by the band structure, the occupation of the electron 
system and the coupling strength.\\
In this work we wish to investigate the energetic structure of the
magnons. The magnitude of the dispersion is influenced by the real part of the magnon 
self energy, while the magnon lifetime is described by its imaginary part. 
The self energy $M_{\bf q}(E)$ can be defined by the formal solution of the 
retarded magnon Green's function 

\begin{equation}
\label{Gq2}
G_{\bf q}^{ret}(E)=\frac{1}{E-M_{\bf q}(E)+i0^+}. 
\end{equation}
The magnon self energy in RPA can now be derived from Eq.(\ref{Gq}):

\begin{equation}
M_{\bf q}(E)=J\langle \sigma^z\rangle+J^2S(Re\chi({\bf q},E)+
iIm\chi({\bf q},E)).
\end{equation}
By applying the Dirac identity we can calculate the real and imaginary part 
of $\chi({\bf q},E)$:

\begin{equation}
\label{Rechi}
Re\chi({\bf q},E)=\frac{1}{2N}{\cal P}\sum_{\bf k}\frac{(\langle n_{{\bf k}\uparrow}\rangle-\langle n_{{\bf k+q}\downarrow}
\rangle)}{E-\epsilon({\bf k+q})+\epsilon({\bf k})-JS},
\end{equation}

\begin{equation}
\label{Imchi}
Im\chi({\bf q},E)=\frac{-\pi}{2N}\sum_{\bf k}(\langle n_{{\bf k}\uparrow}\rangle-\langle n_{{\bf k+q}\downarrow}
\rangle)\delta(E-\epsilon({\bf k+q})+\epsilon({\bf k})-JS)
\end{equation}
where ${\cal P}$ denotes the principal value. This gives rise to two 
collective excitation modes and a one particle continuum as discussed in  
\cite{Babc81}.\\ 
Outside this continuum the principal value at the right-hand side of 
eq.~(\ref{Rechi}) is equal to the function itself,
the imaginary part vanishes and the magnon lifetime becomes infinite. The 
poles of the Green's function, i.e.~the collective excitation
energies, are then defined by the implicit equation

\begin{equation}
\label{magnen}
E-J\langle \sigma^z\rangle-J^2S Re\chi({\bf
  q},E)\stackrel{!}{=}0.
\end{equation}
The stability of the local moment ferromagnet 
at low temperature is determined by the collective excitations from the 
ferromagnetic ground state. The excitations into the acoustic mode are 
certainly dominant. We therefore limit ourselves to an analysis of the 
dependence of the acoustic mode on $J$ and $n$. \\
In order to solve Eq.(\ref{magnen}) an expression for the
$\langle n_{\bf{k}\sigma}\rangle$ is required. It is very common to treat the
electronic system in a
mean-field approximation (see e.~g.~\cite{Babc81}). This leads to a 
situation in which the  electron-magnon interaction is included in the 
investigation of the magnon system, but not of the electronic system. In this 
work we wish to investigate how the ferromagnetic properties of our model 
system are determined by the RPA theory. We therefore limit ourselves to a 
mean-field treatment of the electronic part. In this approximation the 
expectation values of the ${\bf k}$-dependent occupation numbers in 
Eq.~(\ref{Rechi}) is

\begin{equation}
\label{nkmf}
\langle n_{{\bf k}\sigma}\rangle=\frac{1}{e^{\beta(\epsilon({\bf
k})-\mu-\frac{1}{2}Jz_{\sigma}S)}+1}.
\end{equation}
In this work we assume the system to be a simple cubic lattice with 
nearest-neighbor interaction only. The electronic dispersion of such a system 
is given by

\begin{equation}
\label{scdisp} 
\epsilon({\bf k})=-2t\sum_{i=1}^d cos(k_i)+t_0,\:k_i\:\epsilon[-\pi;\pi],\:
t>0.
\end{equation} 
The midpoint of the band $t_0$ is chosen to be zero. The relation between the 
band width $W$ and the hopping term $t$ is $W=12t$.

\section{Results and Discussion}  

In this section we discuss the properties of the spin-wave
dispersion Eq.(\ref{magnen}) with increasing band occupation $n$ at different 
values of the coupling $J$. The coupling of the conduction electrons to the 
local moments leads to an indirect interaction between the spins.\\
In Fig.~\ref{magstruc} spin wave spectra in the [100] direction are sketched 
for $J=0.2 eV$ and $J=0.25 eV$ at different band occupations $n$. 
By filling the conduction band with
electrons the magnon energy increases up to a maximum at a given band
occupation $n$. This maximum value increases with increasing $J$. After
reaching the maximum value close to the quarter-filled band,
the magnon energy decreases monotonically with further increasing band 
occupation.\\
The excitation energy of a spin-wave mode is the energy needed to
cause a deviation of the spin system of unity from the ferromagnetically
saturated state. Hence the larger the spin-wave energy the larger the
ferromagnetic stability. This is of course a qualitative statement which 
must be substantiated by quantitative calculation of, for example, the 
critical temperature $T_c$.\\
In Fig.~\ref{magstruc2} we have sketched the behaviour of the magnon 
energies by further increasing the band occupation. The spin-wave
spectra become negative.\\
The existence of negative excitation energies is of course unphysical
since the magnons are  excitations from the ground state. There is thus
no stable ferromagnetic ground state beyond a critical band occupation
$n_c$. The determination of $n_c$ requires the definition of a stability 
condition. The system becomes ferromagnetically unstable if the magnon
energy becomes negative.\\ 
A widely used parameter for the ferromagnetic stability is the spin-wave 
stiffness $D$. In \cite{Wang97} it has been shown that the expression for
the magnon energy can be rewritten as $E({\bf q})=D{\bf q}^2$ for small
$q$. When $D$ becomes zero the system becomes ferromagnetically
unstable. In the intermediate coupling regime $0.2eV<J<0.7 eV$, this is
indeed a unique criterion since the spin-wave energy becomes zero at 
small $q$.\\However,  an interesting effect can be observed at very small $J$. In 
Fig.~\ref{magstruc3} the magnon spectra for a system with $J=0.1$ eV are
sketched. At $n=0.24$ the spin-wave dispersion starts to decrease
significantly for large $q$ while the spin-wave stiffness remains
almost the same. This effect is known as 'anomalous softening'. A second 
minimum appears in the dispersion curve at the righthand edge of the Brillouin 
zone. At $n=0.255$ the spin-wave energy at the $X$-point becomes
zero. Hence at small $J$ a ferromagnetic instability exists at large
$q$.\\
This effect has been observed experimentally in the manganite
Pr$_{0.63}$Sr$_{0.37}$Mn0$_3$ \cite{Hwang98}. But in
manganite systems a strong coupling between the conduction electrons and 
the localised moments exists. It is thus not very clear, whether or not
the anomalous
softening found in our calculation is comparable to the experimental
results. A more complex theory which assumes some {\textit d}-character of the
conduction electrons has already been presented in order to explain the
measured anomalous softening \cite{Khali99}. A reasonably good agreement 
between theory and experimental data has been found. Despite this 
unsolved problem anomalous softening is certainly a model property of the 
FKLM as well. A very significant consequence of anomalous softening
is the fact that the spin-wave stiffness $D$ is almost constant while the
magnon energy at large $q$ is lowered. The decrease in magnon
energy is equivalent to a decrease in ferromagnetic stability and
hence of the critical temperature $T_c$. In our theory these facts would
lead to an
increase of the relation $\frac{D}{T_c}$. This effect has also been
observed experimentally in manganite systems \cite{Fern98}.
In this paper a ferromagnetic ground state is assumed. It is thus only 
possible to report a unique ferromagnetic instability at  $J=0.1$ eV. 
The cause of the anomalous softening at low $J$ and hence the nature of the 
new phase is currently the subject of further research.\\
We have defined above the critical band occupation $n_c$ where the spin-wave 
stiffness becomes zero and the system ferromagnetically
unstable. By calculating $n_c$ at different coupling $J$ we can sketch the 
ferromagnetic phase in a $J-n$ diagram, shown in Fig.~\ref{pd}.\\
The ferromagnetic phase occupies a large part of the phase diagram. The
critical band occupation increases by increasing the coupling strength
$J$. There is no ferromagnetism around half filling $(n=1)$. One gets a
qualitative explanation of this result by looking at the energy
of different spin states of the conduction electrons at half filling. 
All lattice points are occupied
with one electron. If all electrons are in the spin-up state, no virtual
hopping between the lattice points is possible. Virtual hopping reduces 
the energy of the system. Thus the stable spin configuration corresponds 
to the state with the maximal hopping amplitude, i.e. where all nearest 
neighbour electrons have opposite spin. Since we assume a ferromagnetic
coupling between the two electron subsystems, such
a spin configuration corresponds to an antiferromagnetic order of the 
local moments.\\ 
At $J=0.1$ eV we have sketched two points. The black point represents the band
occupation where the magnon energy vanishes at the right-hand edge of the
Brillouin zone. The second point is sketched at $n_c$. Hence the
procedure described to determine the phase border is no longer
unique.\\
The structure of the ferromagnetic phase of our system is very similar
to those calculated in \cite{Dago98} and \cite{Nagai99}. These
calculations were performed using a variety of different methods, such as 
quantum Monte Carlo calculations, for systems
with different dimensions d and total spins $S$. It is 
therefore quite problematic to compare these phase diagrams
directly. However the similarity between these results is 
remarkable. The qualitative behaviour of the ferromagnetic phase seems
to be a general property of the FKLM.\\    
Finally we have calculated the critical temperature using an approximate
formula which can be derived from the spin-wave approximation of the
Heisenberg model (see e.~g.~\cite{quantmag2}):
\begin{equation}
\label{tce}
k_BT_c=\left( \frac{1}{NS}\sum_{\bf q}\frac{1}{E({\bf q})}\right)^{-1},
\end{equation} 
where $E({\bf q})$ is the spin-wave dispersion of a local moment system
with direct exchange interaction.\\ 
This formula extrapolates the critical temperature from the magnon
energy of the system in the spin-wave regime. A quantitatively better
calculation of the critical temperature would require a self-consistent 
calculation.\\
At this stage we can only give a qualitative description of the
dependence of the critical temperature on the band occupation $n$ and the
coupling strength $J$. In our system the exchange interaction is mediated
by the conduction electrons. The FKLM-Hamiltonian  can be mapped onto an
effective Heisenberg-Hamiltonian \cite{Rex96}. Information about the kind of
interaction is then buried in the effective exchange integrals $J({\bf
  q})$, which are functionals of the electronic self energy. This
description leads to the modified RKKY-interaction \cite{Rex96}. Following
the same idea we can describe the excitation of the local spins in the
spin-wave approximation as a system of noninteracting magnons:

\begin{equation}
\label{tcesf}
H=E_0+\sum_{\bf q}\Omega_{eff}^{RPA}({\bf q})b_{\bf q}^+b_{\bf q}.
\end{equation} 
E({\bf q})=$\Omega_{eff}^{RPA}({\bf q})$ is the spin-wave
dispersion (\ref{magnen}). The renormalised spin-wave energies are
equivalent to the effective exchange integrals in the modified
RKKY-interaction \cite{Rex96}. In our theory we can therefore rewrite the 
approximate formula (\ref{tce}) for the critical temperature.

\begin{equation}
\label{tce2}
k_BT_c=\left( {\frac{1}{NS}}\sum_{\bf q}\frac{1}{\Omega_{eff}^{RPA}({\bf q})}\right)^{-1}
\end{equation}
This equation is used to calculate the dependence of the critical
temperature by evaluating the magnon energies in the whole Brillouin
zone for given $n$ and $J$.\\
Plots of $T_c$ versus $n$  for different $J$ are
sketched in Fig.~\ref{tcn}. From Eq.~(\ref{tce2}) it is evident that the
functional dependence of $T_c$ is 
equivalent to the $n$-dependence of the magnon energy. The critical
temperature increases by increasing the band occupation up to a
maximum. After passing this maximal value the critical temperature
decreases monotonically before it vanishes at the critical band
occupation $n_c$ (black points). Since the error of the numerical
evaluation of $T_c$ becomes very large near the critical band occupation,
the value of $n_c$ was taken from the phase diagram in Fig.~\ref{pd}. The 
good agreement
between the $T_c$ versus $n$ graphs and the respective $n_c$ values supports 
the stability condition using the spin-wave stiffness $D$ defined above.  
Good qualitatitive agreement with the $T_c$ dependence on $n$ reported 
in \cite{Rex96} has been found. This agreement is perhaps unexpected, 
since a more realistic description of the electronic system is used in
\cite{Rex96}.
This could be due to the fact that the exchange integrals used in \cite{Rex96}
are to first order very similar to the expression of the real part of 
$\chi({\bf q}, E)$ in Eq.(\ref{Rechi}) which renormalises the expression of 
the magnon energy Eq.~(\ref{magnen}) with respect to electron-magnon 
interaction. The 
qualitative behaviour of the critical temperature is mainly determined by 
these interaction processes. It is thus reasonable that the inclusion of 
those processes in the perturbative analysis of the magnon system leads 
to a good description of the qualitative behaviour of $T_c$.\\
An exception is the $T_c(n)$ dependence at $J=0.1$eV.  The
calculation stopped at $n=0.21$. Above this band occupation
we have to sum over two
singularites in Eq.(\ref{tce2}) due to the effect of anomalous softening. A
numerical calculation is now rather complicated and was not done at this 
stage. However we sketched the critical band occupation $n_c=0.38$, where
the spin-wave dispersion at small values of $q$ vanishes. If we
assume the same qualitative dependence of $T_c$ as found for the larger
$J$ this $n_c$ is too large. Hence the stability condition using $D$ does not 
work at this $J$.\\
A similar qualitative dependence  of the critical temperature, 
on the band occupation, including a maximum at quarter filling, has also been 
found for the strong coupling limit ($J \rightarrow \infty)$ of the FKLM 
assuming classical spin ($S=1$) and using Monte-Carlo methods \cite{Yuno98}.
\\
Another interesting result concerns the critical temperature as a function 
of the coupling strength $J$ in Fig.~\ref{tcj}. It is sketched for three 
different
band occupations $n$. The $T_c$ versus $J$ dependences for $n=0.1$ and $n=0.06$
are quite similar. The critical temperature increases with increasing
coupling strength $J$. This result is intuitively correct. For $n=0.4$ we
found a critical coupling strength $J_c$. Below $J_c$, $T_c$ is zero and
the system is not ferromagnetically ordered. The existence of a $J_c$
has also been shown in \cite{Rex96} for higher band occupations. A
different qualitative behaviour compared to \cite{Rex96} has been found
at large $J$. While in that work $J$ remains constant above a certain
value, we did not find saturation. This difference could be caused by
the different electronic self energies used. We have not calculated the 
critical temperature for $J<0.1$eV. For this region we expect the normal 
RKKY-behaviour $T_c\sim J^2$, as shown in \cite{Rex96}.\\
   
\section{Summary and Conclusions}
We have investigated the ferromagnetic properties of the FKLM through an 
analysis of the spin-wave spectrum at very low temperature. The spin-wave 
approximation was applied to the model Hamiltonian and the magnon 
Green's function was derived within the RPA approximation. The numerical 
analysis of the magnon energy spectrum was limited to the acoustic branch of 
the collective excitations.\\
The results of $n$- and $J$-dependent calculations of the magnon spectra
have been presented. The magnon energy for fixed coupling strength
 $J$ increases with increasing occupation of the conduction band up to 
a maximum value close to quarter filling. After reaching the maximum value the
energy decreases as n increases and eventually vanishes. 
A stability
condition for the ferromagnetic state was defined by using the spin-wave 
stiffness $D$. If the magnon energy at small $q$ vanishes at the critical band 
occupation $n_c$, i.e. $D$ becomes zero, the system becomes ferromagnetically 
unstable. The ferromagnetic phase in three dimensions
has been determined by calculating $n_c$ for different $J$. There is a good 
agreement to the ferromagnetic phase obtained for FKLM-model systems with 
different total spin $S$ and dimension $d$ in \cite{Dago98} and 
\cite{Nagai99}. The critical band occupation $n_c$ 
increases by increasing the coupling strength. There is no 
ferromagnetism around $n=1$. The good agreement between the different 
calculations may lead to the conclusion that the structure of the 
ferromagnetic phase is a general property of the FKLM.\\
In addition anomalous softening is found at $J=0.1 eV$. The method to determine
$n_c$ is thus no longer unique. It is not yet possible to link this
result to the anomalous softening observed experimentally. This problem as well
as the physical reason  of this phenomenon is an interesting subject of 
future research.\\
Finally the critical temperature dependence on $J$ and $n$ was presented. A 
qualitative agreement to the results shown in \cite{Rex96} was found. A more 
careful treatment of the electronic system and a 
selfconsistent calculation of the critical temperature is currently
the subject of further research.

\newpage
\begin{figure}[bth]
\newcommand{\myscale}{0.28}
\epsfig{file=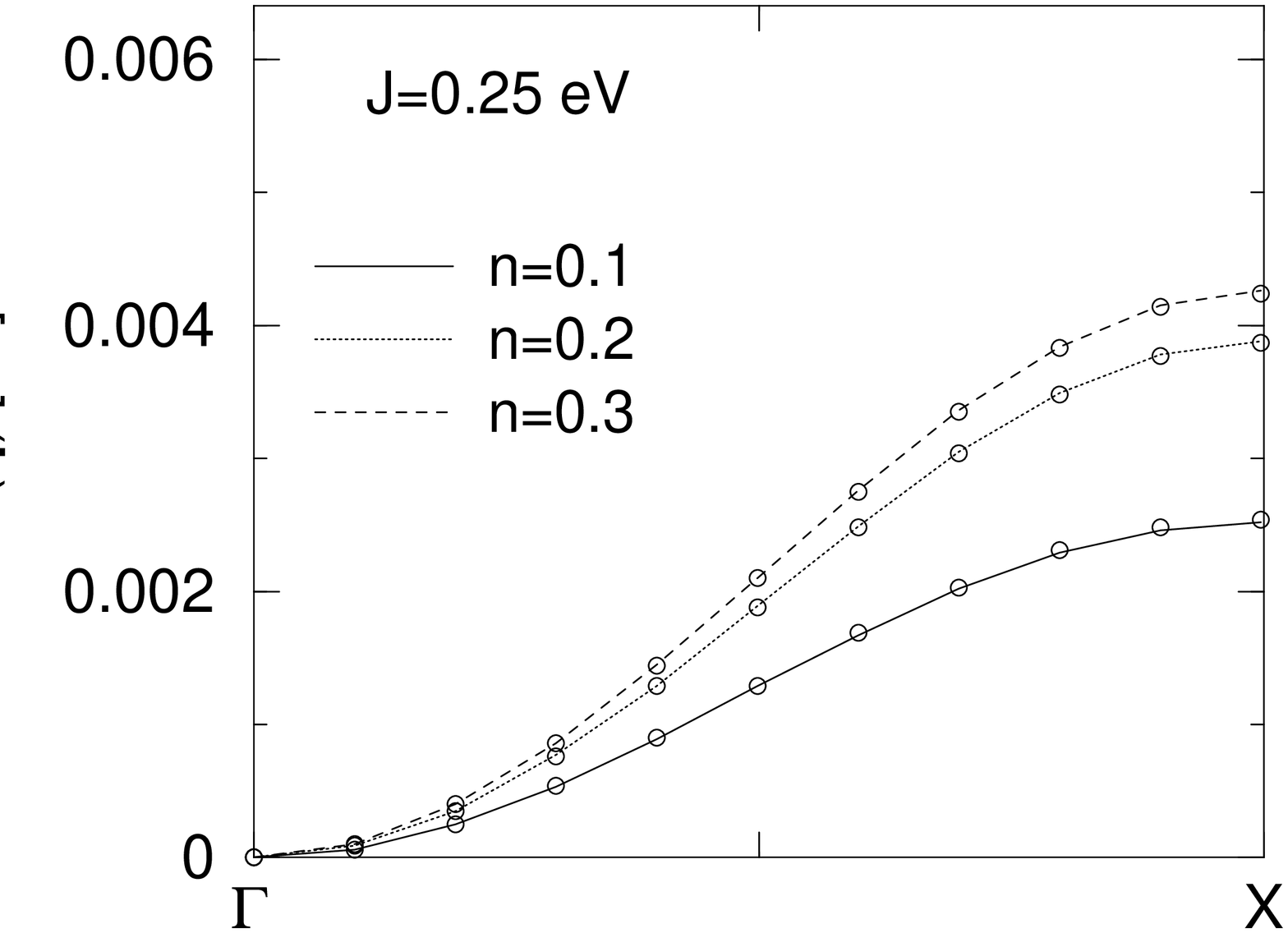,scale=\myscale, angle=270}\hfill
\epsfig{file=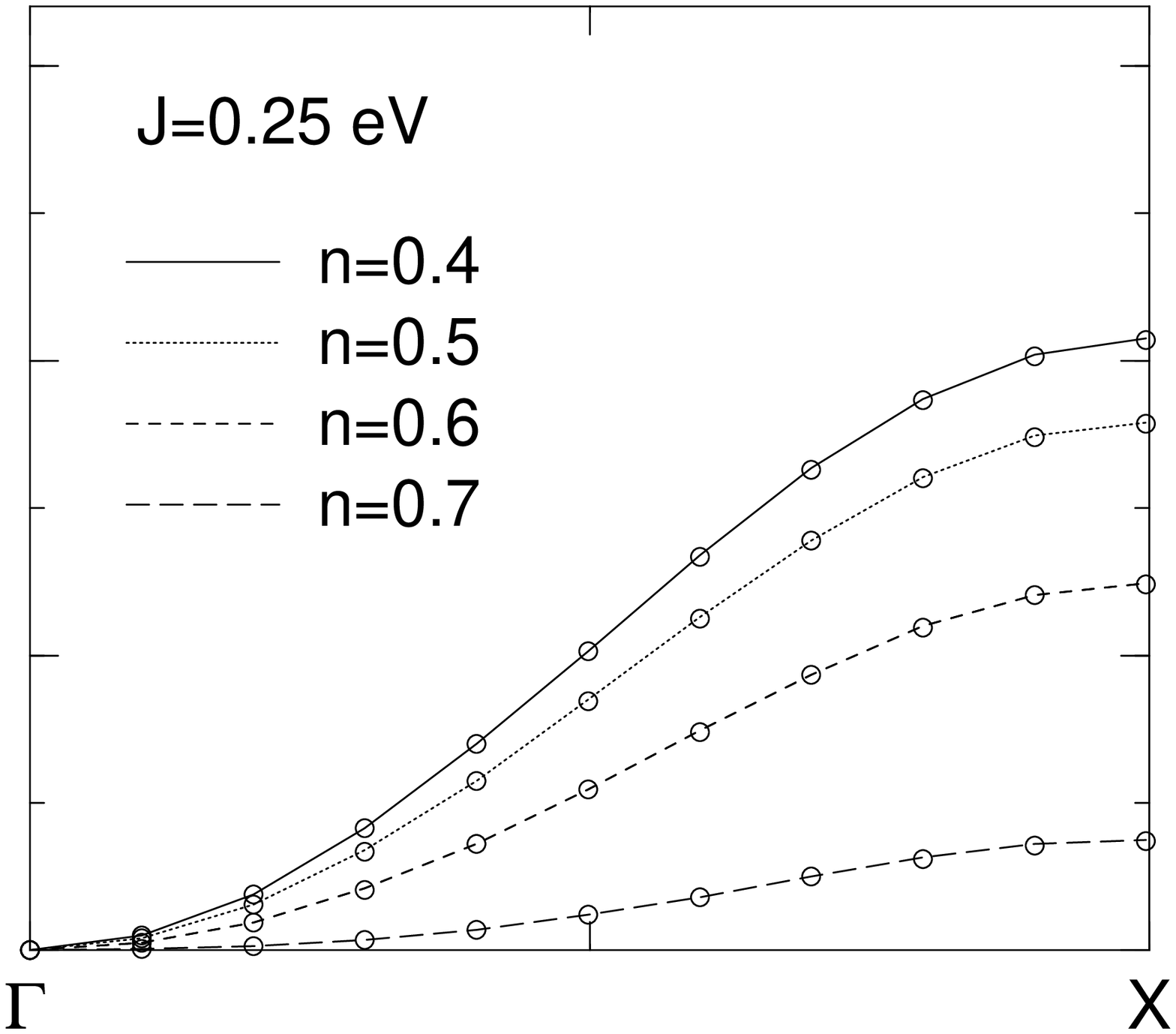,scale=\myscale, angle=270}\\
\epsfig{file=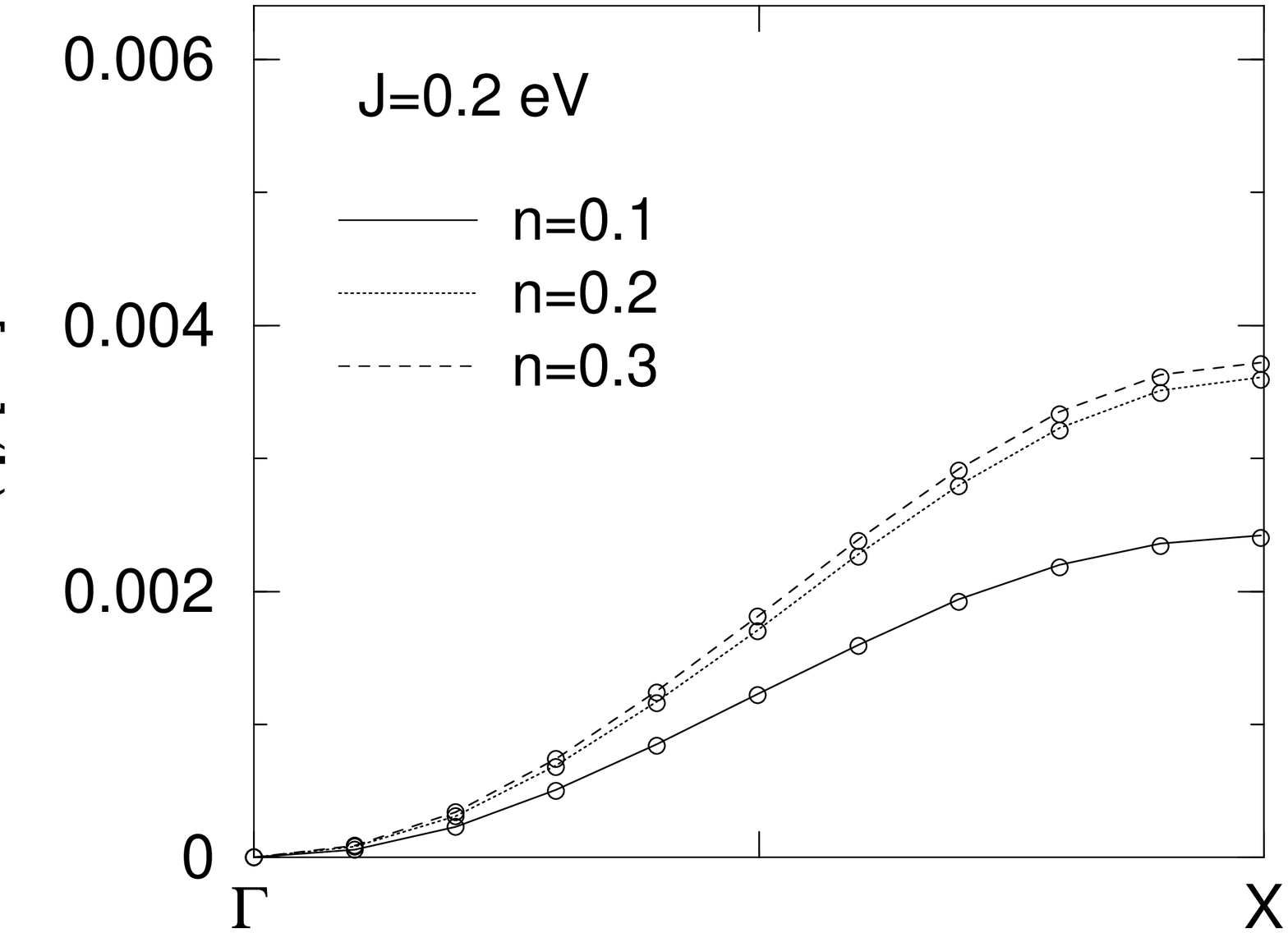, scale=\myscale, angle=270}\hfill
\epsfig{file=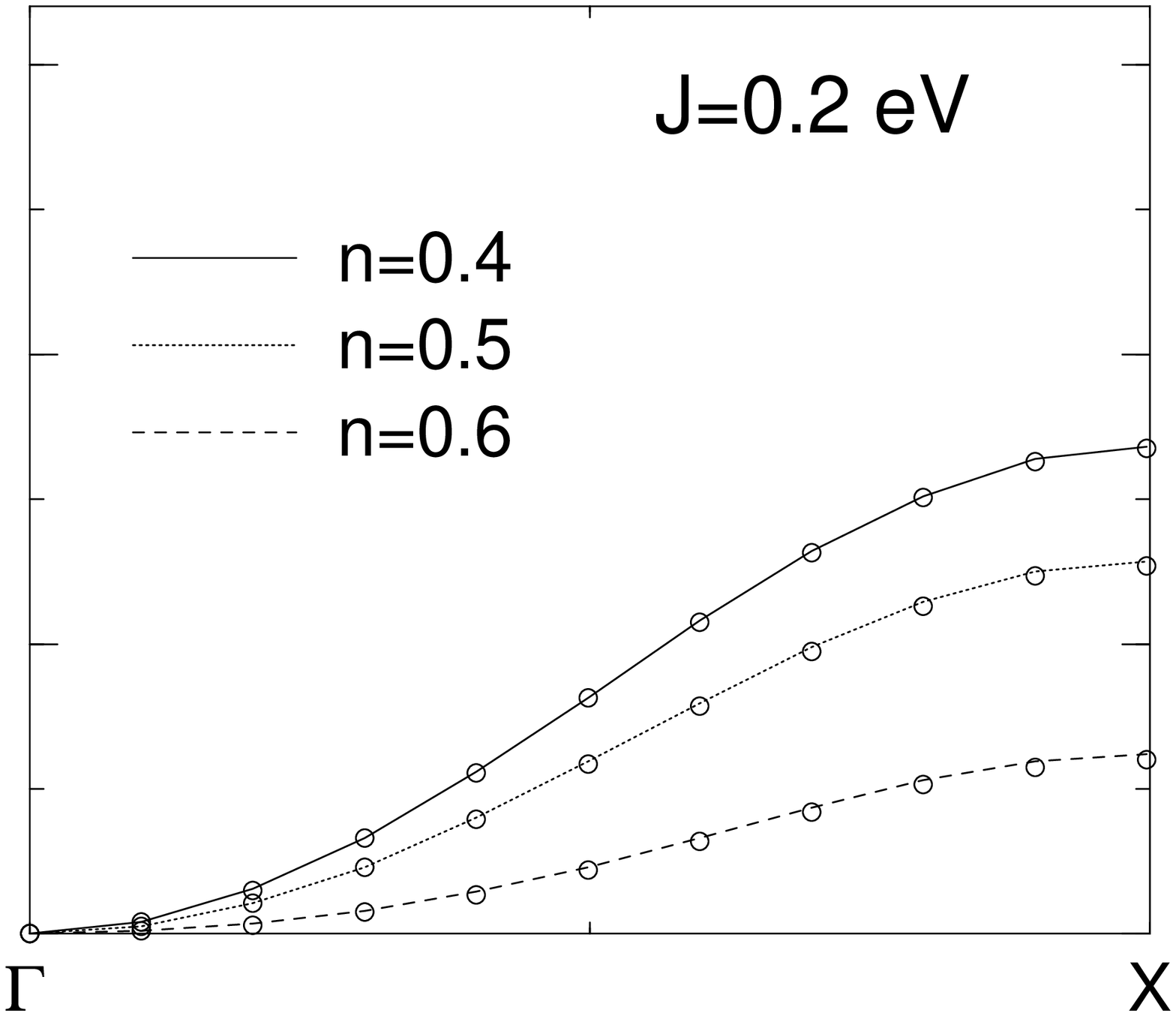, scale=\myscale, angle=270}\\
\caption{\label{magstruc}Spin-wave dispersion $E({\bf q})$ in [100]
direction  for a
  system with $S=3.5$, band width $W=1.0$ eV at $T=0$ K. The dispersions for 
two
  different coupling constants $J=0.2$ eV and $J=0.25$ eV at different
  band occupations $n$ are shown. On the left-hand side of the figure
 the magnon energy increases with increasing n, whilst at the right-hand side 
it decreases.} 
\end{figure}
\begin{figure}[bth]
\newcommand{\myscale}{0.28}
\epsfig{file=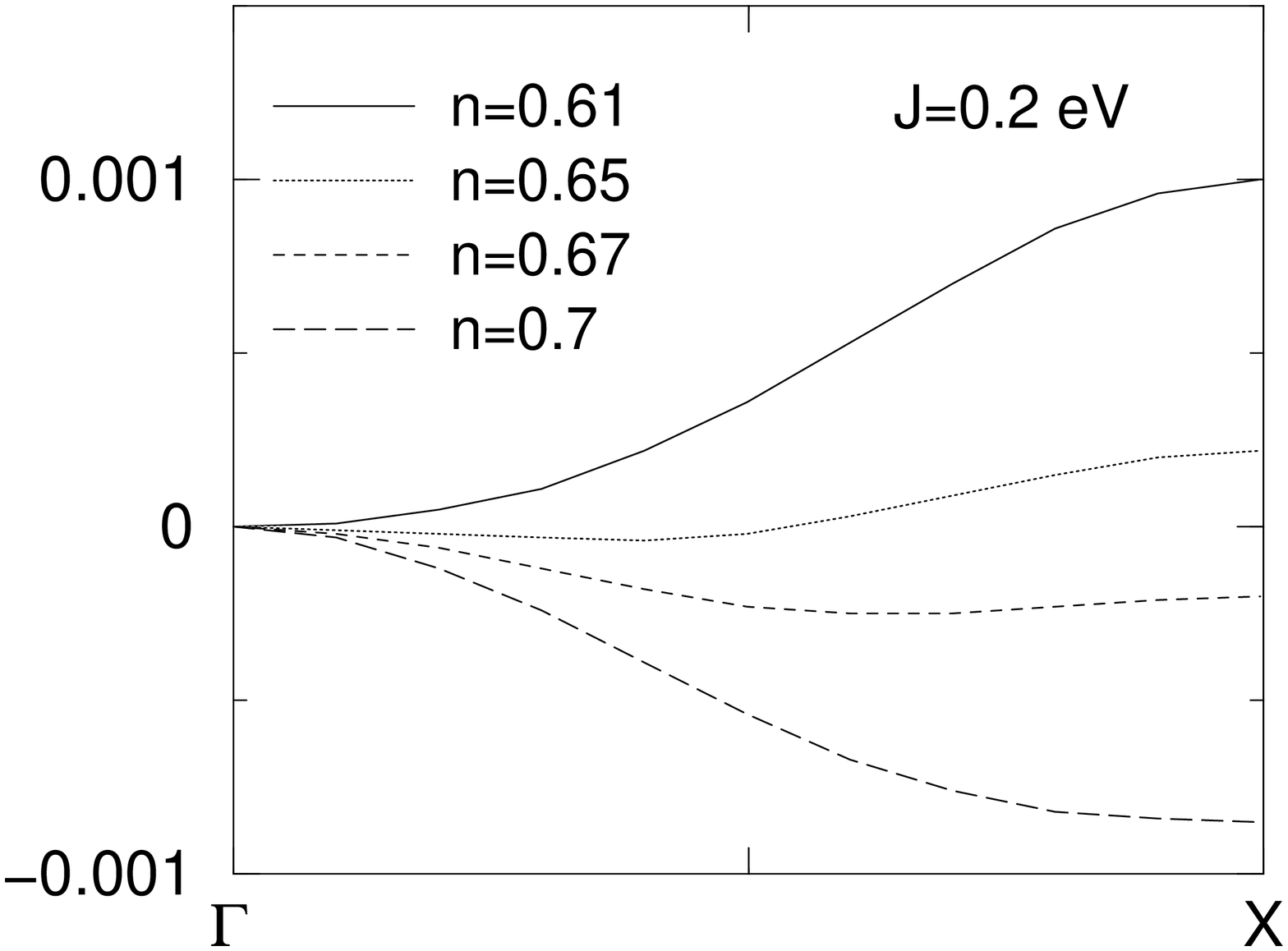, scale=\myscale, angle=270}\hfill
\epsfig{file=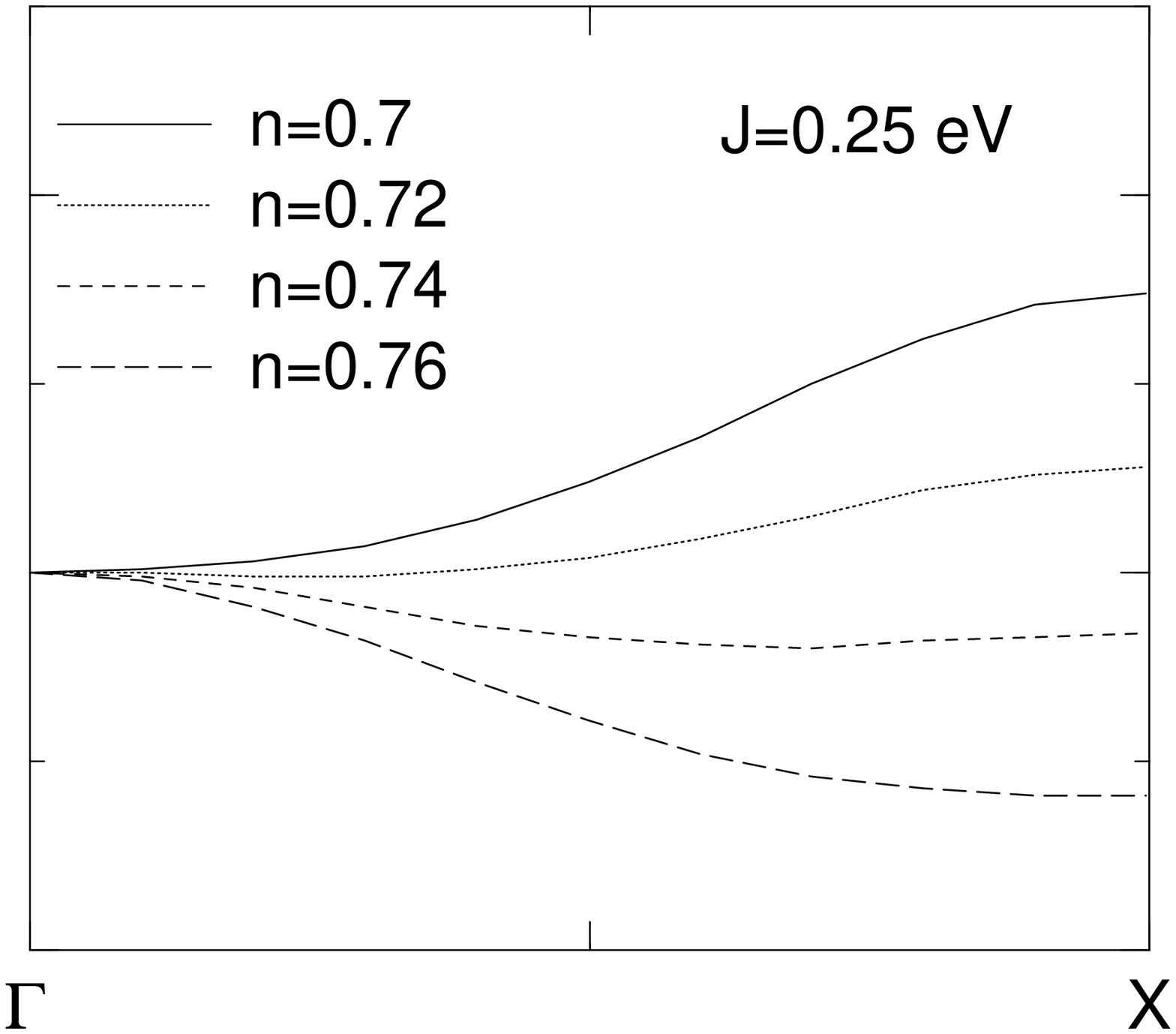, scale=\myscale, angle=270}\\
\caption{\label{magstruc2}Spin-wave dispersion $E({\bf q})$ in [100] 
  direction for a
  system with $S=3.5$, band width $W=1.0$ eV at $T=0$ K. The dispersions for 
 two
  different coupling constants $J=0.2$ eV and $J=0.25$ eV at different
  band occupations $n$ are shown.} 
\end{figure}
\begin{figure}[bth]
\newcommand{\myscale}{0.3}
\epsfig{file=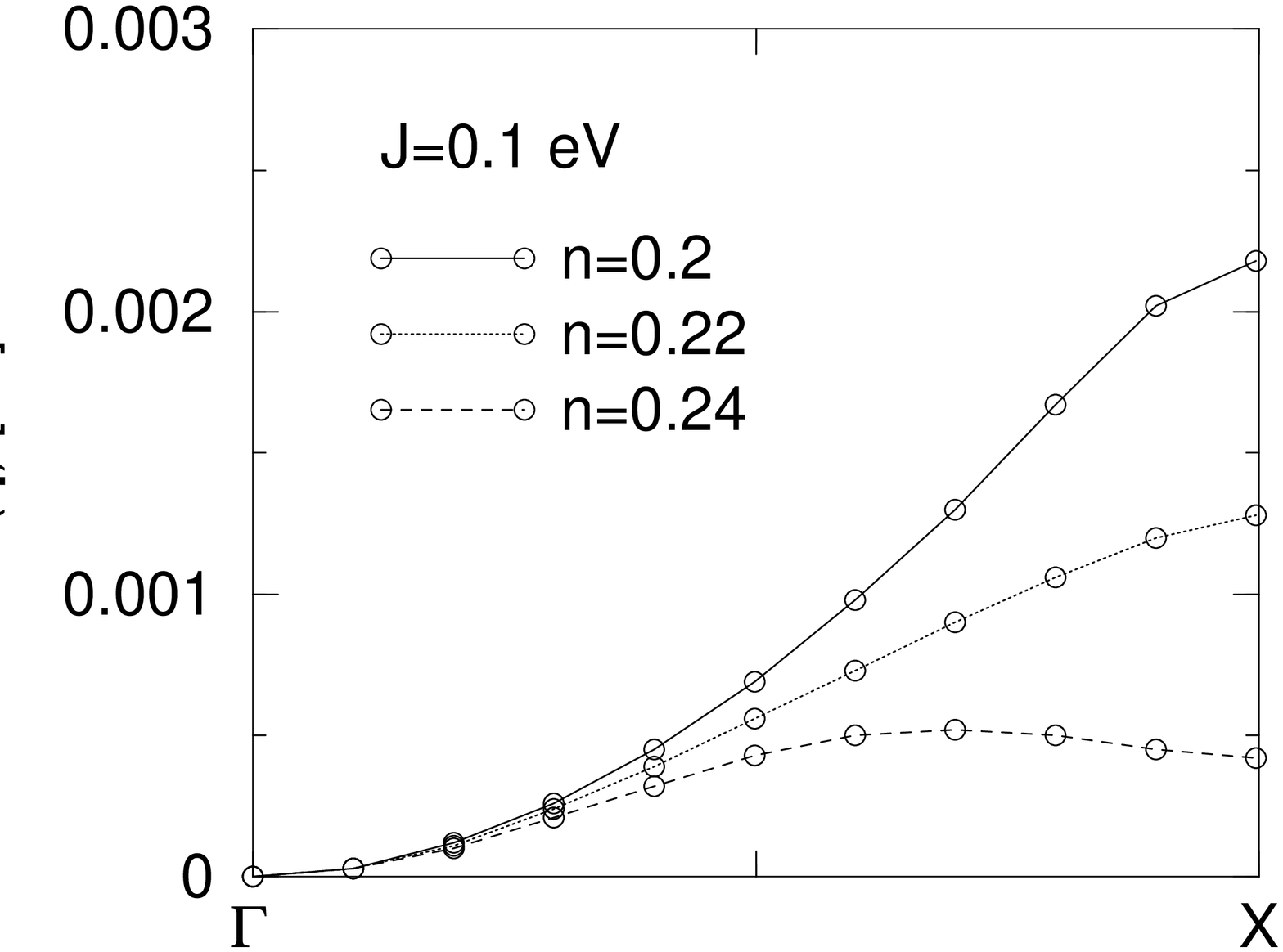, scale=\myscale, angle=270}\hfill
\epsfig{file=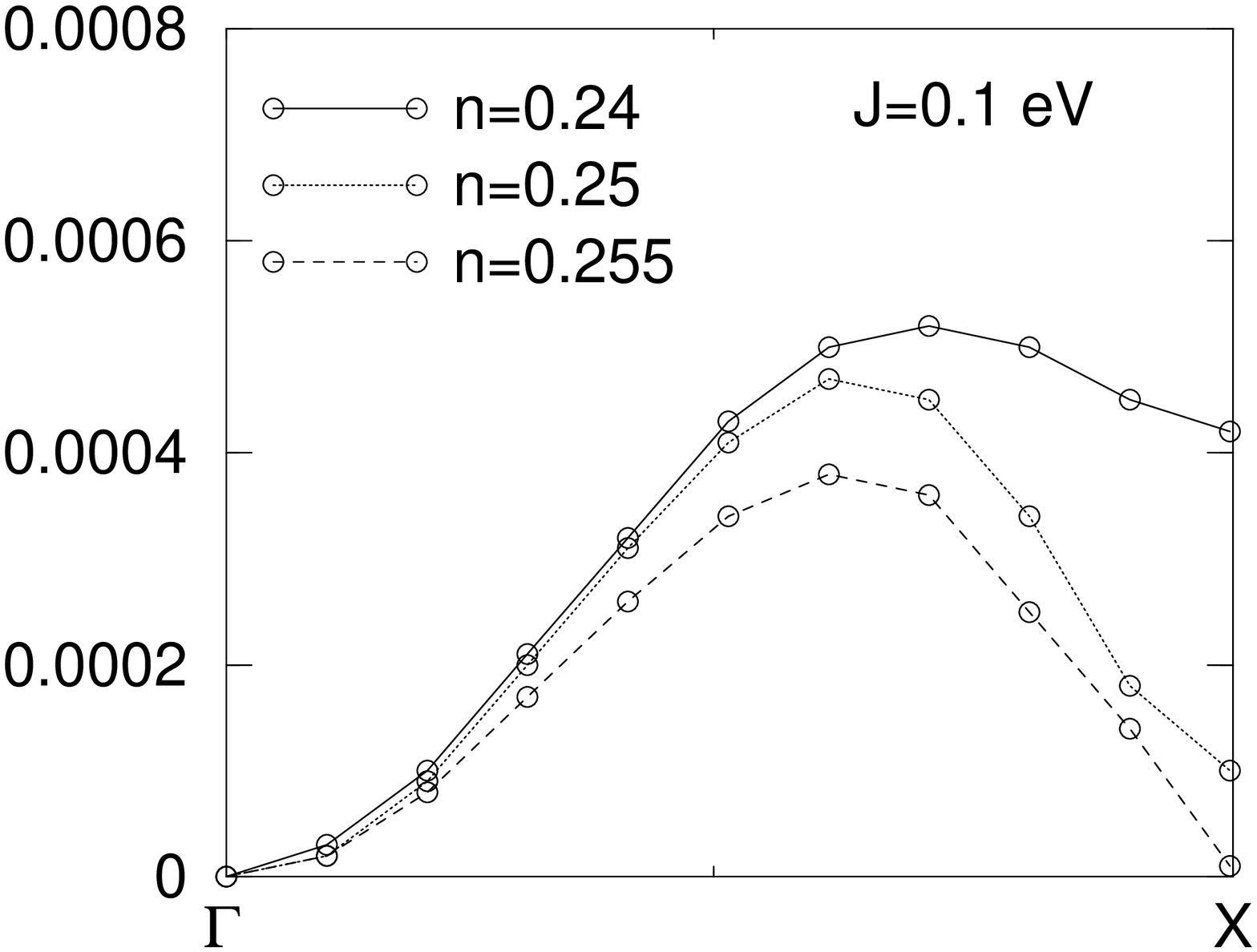, scale=\myscale, angle=270}\\
\caption{{\label{magstruc3}Spin-wave dispersion $E({\bf q})$ in [100]
    direction for a
  system with $S=3.5$, band width $W=1.0$ eV at $T=0$ K. The dispersion for 
different $n$ at $J=0.1$ eV are shown.}} 
\end{figure}
\begin{figure}[bth]
\newcommand{\myscale}{0.4}
\epsfig{file=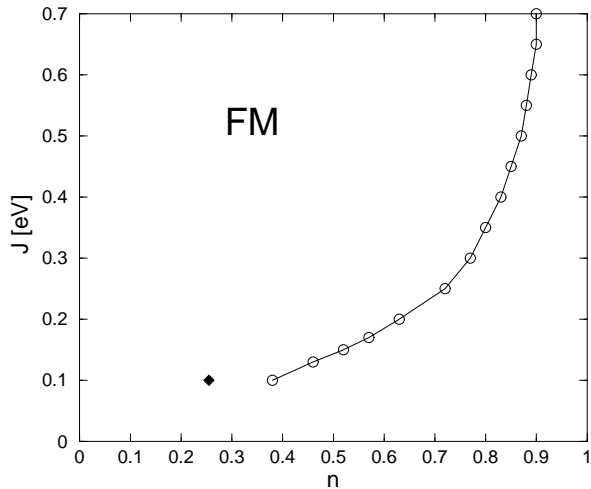, scale=\myscale, angle=270}
\caption{\label{pd}Ferromagnetic phase at $T=0$ K, $S=3.5$, and $W=1.0$ eV} 
\end{figure}
\begin{figure}[t]
\newcommand{\myscale}{0.4}
\epsfig{file=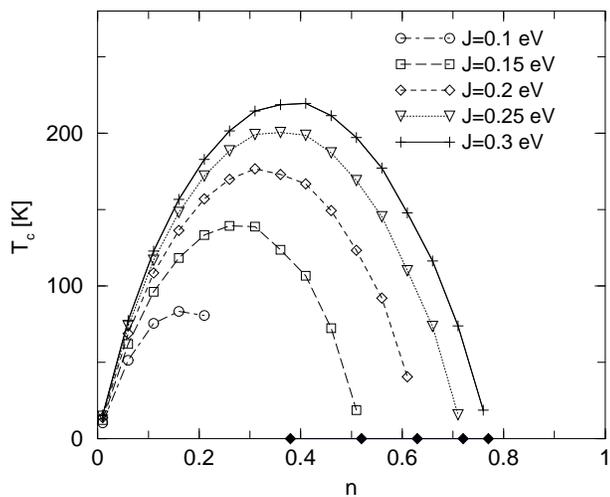, scale=\myscale, angle=270}
\caption{\label{tcn}Critical Temperature $T_c$ as a function of $n$ at 
different coupling constants $J$.} 
\end{figure} 
\begin{figure}[t]
\newcommand{\myscale}{0.4}
\epsfig{file=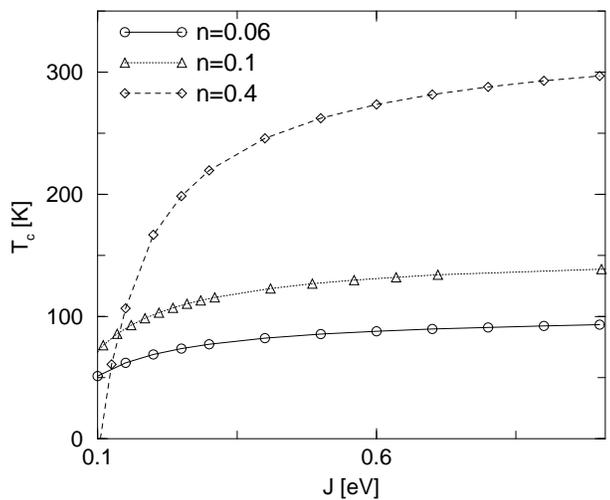, scale=\myscale, angle=270}
\caption{\label{tcj}Critical Temperature $T_c$ as a function of $J$ at 
different band occupations $n = 0.06,~0.1,~0.4$.} 
\end{figure}  
\end{document}